**Different Cost Performance: Different Determinants?**

The Case of Cost Overruns in Dutch Transportation Infrastructure Projects

By

Chantal C. Cantarelli[1], Bert van Wee, Eric J. E. Molin, and Bent Flyvbjerg

[1] Corresponding author

## Abstract

This paper examines three independent explanatory variables and their relation with cost overrun in order to decide whether this is different for Dutch infrastructure projects compared to worldwide findings. The three independent variables are project type (road, rail, and fixed link projects), project size (measured in terms of estimated costs) and the length of the project implementation phase. For Dutch projects, average cost overrun is 10.6% for rail, 18.6% for roads and 21.7% for fixed links. For project size, small Dutch projects have the largest average percentage cost overruns but in terms of total overrun, large projects have a larger share. The length of the implementation phase and especially the length of the pre-construction phase are important determinants of cost overruns in the Netherlands. With each additional year of pre-construction, percentage cost overrun increases by five percentage points. In contrast, the length of the construction phase has hardly any influence on cost overruns. This is an important contribution to current knowledge about cost overruns, because the period in which projects are most prone to cost overruns is narrowed down considerably, at least in the Netherlands. This means that period can be focused on to determine the causes and cures of overruns.

## Introduction

Time and again, even during implementation, prevailing cost increases come to light in large-scale transport infrastructure projects. By the time of opening, the calculated cost overruns are enormous. That these cost overruns are a severe problem has been shown in previous studies (Flyvbjerg et al., 2003a, 2003b; Merewitz, 1973; Pickrell, 1992; Odeck, 2004; Nijkamp and Ubbels, 1999).

From a first study on the cost performance of Dutch projects, it was concluded that the frequency and magnitude of cost overruns is considerably smaller than that found in other studies (Cantarelli et al., forthcoming). Differences between the results of studies can usually be ascribed to one of the following explanations. First, the difference in use of nominal and real prices (Flyvbjerg, 2007). Second, the way in which data is handled; different base for estimated and actual cost (see for a more elaborative description Flyvbjerg et al., 2003b). Third, sample size could be an explanation and fourth the difference in geographical area of project type that is covered.

The Dutch study has followed as much as possible the international research by Flyvbjerg et al. (2003a), it is highly unlikely that differences between both studies can therefore be ascribed to the first three reasons for different cost performance. We conclude that cost overruns in the Netherlands are different from that in other countries, and hence geography matters. To explain the difference in more detail, this research will investigate the determinants for cost overruns in Dutch projects.

Of the various determinants of cost overruns that could be considered in this respect, we will consider the following three in this paper: i.e. project type, project size and the implementation phase. These determinants have been addressed in previous studies into cost overruns and seem to be the most important in understanding cost overruns. Some first expectations about these variables are described below.

Cost performance usually differs between project types, with typically the largest cost overruns incurred for rail projects and more reserved overruns for road projects (Flyvbjerg, 2003b; Merewitz; 1973, Morris, 1990) (with the exception of the findings in the study by the Auditor General of Sweden, found in Odeck, 2004). A larger share of road projects could explain the smaller average cost overrun in Dutch projects (assuming project type matters).

Much less consensus exists with respect to the impact of project size. We define project size, in line with standard convention, in terms of estimated costs. Odeck (2004) reported that "cost overruns appear to be more predominant among smaller projects as compared to larger ones", whereas Flyvbjerg et al. (2004) found that "…the risk of cost escalation is high for all project sizes…".

Compared to project type and project size, the length of the implementation phase is to a lesser extent addressed in previous studies but it turns out to be an essential predictor of cost overruns. According to Flyvbjerg et al. (2004) it is even more vital than project type: "If information on implementation duration is given, project type is not important". Projects with longer implementation phases have larger cost overruns. If Dutch projects would have shorter implementation phases, this may explain why in the Netherlands cost overruns are relatively small.

To summarise, this paper aims to investigate whether project type, project size and the implementation phase are also relevant for the variance in cost overruns in the Netherlands and whether these variables can explain the differences in cost performance between Dutch projects and those found in other studies.

The remainder of this paper is organised as follows: Section 2 will describe the data collecion and methodology. After that, the relation between each of the three determinants on cost overruns is described in sections 3 to 5. Section 6 draws the main conclusions and discussions and finally section 7 describes areas for further research.

## Data Collection and Methodology[1]

### Data Collection

All large-scale transport infrastructure projects in the Netherlands, defined as projects that cost more than 20 million Euros (2010 prices) that were completed after the year 1980 were selected. Projects completed before this year are excluded because the data were expected to be difficult to come by. Data were collected from a variety of sources, amongst others interviews with former project leaders and project teams; archives research at the Ministry of Infrastructure and the Environment; RWS[2] Direction Large Projects and RWS Direction Zuid-Holland; internet search; and the MIRT reports. The MIRT (*Meerjarenprogramma Infrastructuur, Ruimte en Transport*, translated as the Multi-year programme for infrastructure, spatial planning and transport[3]) is the implementation programme related to the policy of 'mobility and water' and includes all infrastructure projects in the Netherlands. A total of 161 transport infrastructure projects were selected (road, rail, tunnels and bridges), however, because of reasons of limited data availability or invalidity of the data, 83 projects could not be included in the database The final database therefore consists of 78 projects.

The resulting database does not include all projects due to incompleteness of information. However, in line with previous international research in this field that also includes projects based on data availability, the database is considered to be a sample. In this research also non-significant differences will be reported because we are also interested in a complete description of the project performance of the specific projects in the database.

The database is considered representative for the population of road and rail projects, but some bias is expected regarding cost overruns for tunnels and bridges. For these projects, data was partly based on interviews, which holds the risks that more favourable data is given. The cost overruns for fixed link projects may therefore be underestimated.

**Methodology**

The two most important data variables in this research are the estimated and actual costs. Cost overrun is measured as actual out-turn costs minus estimated costs expressed as a percentage of the estimated costs. Actual costs are defined as real, accounted construction costs determined at the time of project completion. Estimated costs are defined as budgeted or forecasted construction costs determined at the Time of formal Decision to build (ToD). This is also called the "decision date", "the time of the decision to proceed," the "go-decision" (Flyvbjerg et al., 2003a). At that moment, cost estimates were often available as data for decision-makers to make an informed decision.

By means of statistical analyses we will investigate whether the three determinants have any relation with cost overrun and how this relation can be described.

In line with Flyvbjerg et al. (2004) we define the implementation phase as the period from the year of the formal decision to build (ToD) until the construction is completed and operations have begun. Data about the year in which operations have begun (here referred to as the actual opening year) is therefore required. However, for projects that were based on the MIRT documentation, data on this actual opening year were unavailable and an assumption had to be applied. For projects for which the year of opening is indicated in the MIRT, the assumption was established by comparing the year of opening with the last year in which costs were provided in the MIRT. It turned out that, on average, the actual opening year was one year (road projects) or one and a half years (rail projects) before the last year for which costs were indicated in the MIRT. Resulting from these findings we assume that the actual opening year for road projects is one year and for rail projects one and a half years before the last year for which costs were indicated in the MIRT. These assumptions are considered fairly reasonable since the MIRT is prepared one year before it is published (MIRT of 2004 is set up in September 2003).

The implementation phase is split into two phases: the pre-construction phase and the construction phase. The pre-construction phase is the period between the ToD and the start of construction. The construction phase is the period between the start of construction and the year in which the project is completed and operation has begun. The cost overrun in the preparation phase is measured as the estimated costs at the start of construction minus the estimated costs at the ToD expressed as a percentage of the estimated costs at the ToD. The cost overrun in the construction phase is measured as the actual out-turn costs minus the estimated costs at the start of construction expressed as a percentage of the estimated costs at the start of the construction. For the analyses regarding these two phases, similar to the study in the companion paper (Cantarelli *et al.,* forthcoming) only projects that consist of a preparation and construction phase are included.

## Cost Overruns per Project Type

For each project type this section presents information on cost overruns including the average cost overrun and the frequency with which cost overruns occur in general, and by two different project phases specifically (the pre-construction and the construction phase).

### Characteristics of Cost Overruns per Project Type

Table 1 gives an overview of the average cost overruns for each project type.

**Table 1 Average cost overrun per project type**

| Project Type | N | Mean CO % | SD |
|---|---|---|---|
| Road | 37 | 18.6 | 38.9 |
| Rail | 26 | 10.6 | 32.2 |
| Fixed links | 15 | 21.7 | 54.5 |
| Bridges | 7 | 6.5 | 33.3 |
| Tunnels | 8 | 35.0 | 67.4 |
| **Total** | 78 | 16.5 | 40.0 |

Fixed link projects have the largest average cost overrun of 21.7%, followed by road projects with 18.6% and rail projects with 10.6% (F=0.458, p=0.634). Subdividing fixed links into bridges and tunnels, we find that tunnels appear to be considerably more prone to cost overruns than bridges though the difference is not significant (F=1.021, p=0.331) – note that the numbers are low. It should be noted that the presence of lock-in is highly likely and hence these cost overruns are underestimated (see Cantarelli *et al.* forthcoming for an elaboration of the increase of the average cost overrun due to lock-in).

A possible explanation for the low average cost overrun for rail projects is the type of construction. It is possible that the rail projects included in this research are mostly expansions of existing railway lines e.g. broadenings (from two tracks to four tracks), improvements or adjustments, rather than new infrastructure constructions. These types of constructions usually involve smaller cost overruns than new infrastructure. For the Dutch data the average cost overrun for the projects (road, rail, fixed links) that concerned expansions of existing infrastructure was indeed 9.2% lower (SD=34.0) than the average of 20.9% (SD=42.9) for the projects concerned with the construction of new infrastructure (t=1.256, p=0.213, independent sample t-test). This does not only apply for all projects together but also for road and rail projects separately. However, the share of projects which expand existing infrastructure in rail projects is not higher than for road projects. The type of construction cannot therefore explain the difference in average cost overrun between these project types. Organisational set-up and institutional settings may account for the difference between the project types – ProRail is project owner for rail projects and RWS for road projects.

The relatively low cost overruns of Dutch rail projects compared to worldwide rail projects may be explained by the type of rail projects. Dutch rail projects mainly concern heavy rail whereas the worldwide research also concerns light rail, a type of rail that typically involves much higher cost overruns.

As the number of tunnel and bridge projects is considerably smaller compared to the number of road or rail projects and since analyses based on a small number of projects are much more vulnerable to extreme scores, tunnels and bridges are taken as one category called fixed links in the remainder of this paper. The subdivision will thus be based on three project types: road, rail and fixed link projects.

Table 2 presents the frequency by which cost underruns (left side) and cost overruns (right side) occur.

**Table 2 Number of projects with cost underrun and overrun (in percentage and number) and their averages**

| Project type | Number of projects with cost underrun | | Number of projects with cost overrun | | Mean Cost underrun % (SD) | Mean Cost overrun % (SD) |
|---|---|---|---|---|---|---|
| | (%) | (#) | (%) | (#) | | |
| *Road* | 37.8 | 14 | 62.2 | 23 | 14.3 (12.7) | 38.7 (35.6) |
| *Rail* | 50.0 | 13 | 50.0 | 13 | 13.1 (8.8) | 34.2 (29.4) |
| *Fixed links* | 53.3 | 8 | 46.7 | 7 | 14.4 (10.1) | 62.9 (55.5) |
| ***Total*** | 44.9 | 35 | 55.1 | 43 | 13.9 (10.5) | 41.3 (38.1) |

For road projects, more projects have cost overruns (62%) than cost underruns (38%) (p=0.188 respectively, binomial test). Furthermore, the magnitude of overruns (the mean overrun is 39%) is also higher than that of the magnitude of underruns (14%) (p=0.009). For rail and fixed link projects, the frequency of overruns is equal to or smaller than the frequency of underruns. However, for both rail and fixed projects, again, the magnitudes of overruns are higher than that the magnitudes of underruns (p=0.006 and p=0.011 respectively). The average cost underrun is similar between project types (p=0.948, Anova) but the average overrun is twice as large for fixed link projects compared to road and rail projects (p=0.249). Remarkably, fixed links have the lowest frequency of cost overruns, but the average cost overrun is largest.

**Cost Overruns in the Pre-Construction and Construction Phase**

The main problem with cost overruns lies in the pre-construction phase but this does not necessarily mean that this also applies for each project type[2].

Table 3 indicates the cost overruns in the pre-construction and in the construction phase for each project type specifically. It presents the frequency by which cost underruns and cost overruns occur and the respective average underrun and overrun. In the following we will focus on the results for road and rail projects and not on fixed link projects because the number of fixed link projects is too small to make any statements.

Table 3 Average cost underrun and overrun in the pre-construction and construction phase per project type[b]

| Project Type | N | Pre-construction phase | | | | Construction phase | | | |
|---|---|---|---|---|---|---|---|---|---|
| | | Mean | SD | # projects underrun-overrun | % underrun-overrun | Mean | SD | # projects underrun-overrun | % underrun-overrun |
| *Road* | 23 | 17.6 | 33.5 | 5-18 | 12 - 26 | -2.9 | 15.2 | 12-11 | 14 - 9 |
| *Rail* | 11 | 21.5 | 33.1 | 5-6 | 2 - 41 | -6.9 | 14.2 | 9-2 | 12 - 16 |

[b] In which CU= cost underrun and CO=cost overrun and #=number of

Considering the pre-construction phase, the main findings are as follows:

- The average cost overrun is smallest for road projects and largest for rail projects (p=0.098, F=0.756, One way Anova).
- Cost overruns are more common than cost underruns for both road and rail projects (p=0.011 and p=1.000 for road and rail respectively, binominal test).
- For projects with cost underruns, rail projects have the largest underrun (12%; F=2.246, p=0.172, One way Anova). For projects with cost overruns, rail projects have the largest average cost overrun (41%), followed by road projects (26%) (F=0.936, p=0.344, One way Anova).

---

[2] Note that actual cost overruns can only take place once the project is completed and payments are made, but for simplicity we stick with this term.

Considering the construction phase, the main findings are as follows:
- On average, road and rail projects involve cost underruns of 3% and 7% respectively. (F=0.545, p=0.466, One-way Anova).
- Cost underruns are more common than cost overruns (p>0.05 for both road and rail, binominal tests).
- For projects with cost overruns, rail projects have the largest average cost overrun (F=1.581, p=0.235, One way Anova). For projects with cost underruns, road projects have the largest underrun (F=0.117, p=0.736, One way Anova).

To summarise, in the pre-construction phase most road and rail projects had cost overruns and in the construction phase most projects had cost underruns. The average cost overrun in the pre-construction phase is significantly higher than the average cost underrun in the construction phase (p=0.011 and p=0.034 for road and rail projects respectively, Paired-sample T-test). It appears that the cost performances in both phases are of a different nature. Moreover the project types perform differently in both phases. Though we did not investigate the mechanisms involved, we assume that before the construction starts, projects tend to get more expensive to correct for optimism at the decision to build, or because of additional costs due to additional measures taken to reduce the impact on the environment. Once the construction phase has started, cost overruns are less common and a lot of projects are built for the costs as estimated at the start of the construction phase. Obviously large cost overruns in this phase due to strategic behaviour (having resulted in too low cost estimates at the moment the construction starts) or due to unforeseen (or at least not considered) technical problems are less common.

## Project Size

Project size will be examined in this study in two ways; as an ordinal and scale variable. First of all, projects are often categorised as small, medium, large or very large projects and then the differences in the average percentage cost overrun between these groups of projects is determined. Secondly, as a scale variable, the influence of project size on the extent of the cost overruns is examined. These two subjects will be addressed in sections 4.1 and 4.2 respectively.

Two projects were considered statistical outliers and are excluded from the analyses. Both projects had estimated costs of more than € 4000 million whereas the average project size for the other projects was €147 million (SD=168) (2010 prices). These projects are the Betuweroute and HSL-South, two recently implemented rail projects that are also different from the other projects in the database in terms of their length (160 km and 125 km respectively, compared to the average length of the other projects of 5 km).

**Cost Overruns for Small, Medium, Large and Very Large Projects**

Small, medium, large and very large projects were defined by the cost limits that were used in the MIRT of €112.5 and €225 million. The MIRT (*Meerjarenprogramma Infrastructuur, Ruimte en Transport*, translated as the Multi-year programme for infrastructure, spatial planning and transport[2]) is the implementation programme related to the policy of 'mobility and water', and is part of the budget of the infrastructure fund of the Ministry of Infrastructure and the Environment. These two cost limits result in 3 categories with the category including the smallest projects representing almost 75% of all projects. Because of its high share we split this category into two groups by introducing a third cost limit of €50 million (again half the cost of the first limit of €112.5 million). The distribution of the projects regarding project size is then as follows:

- Small < €50 million: 24
- Medium €50 - <€112.5 million: 22
- Large €112.5 - <€225 million: 13
- Very large > €225 million: 17

Table 4 presents the statistics on cost overruns broken down by project size and project type. The statistics include the number of projects, the percentage of projects, the average percentage overrun and standard deviation and the net total overrun in percentages. The net total overrun is the overrun in million euros expressed as a percentage of the total overrun in million euros.

**Table 4 Cost overruns broken down by project size (estimated costs in € in 2010)[c]**

| Project size | Number of projects | Percentage of projects | Mean cost overrun (%) | Standard deviation | % of cost overrun |
|---|---|---|---|---|---|
| Small | 24 | 31.6 | 18.5 | 40.5 | 6.3 |
| Medium | 22 | 28.9 | 23.2 | 53.2 | 35.0 |
| Large | 13 | 17.1 | 7.0 | 29.3 | 9.0 |
| Very Large | 17 | 22.4 | 10.9 | 26.7 | 49.7 |

[c] In which:, % of cost overrun = the net total overrun as a percentage for the specific category .

The number of projects in each category is rather evenly distributed amongst the categories with slightly more projects in the small and medium groups and slightly less projects in the large and very large groups. Considering the average overrun, small and medium projects have the largest average cost overrun (F=0.565, p=0.640, univariate analysis of variance). Possible differences between the four groups in average cost overruns could be caused by the way in which the formation of the groups was based on the cost limits of the MIRT. If, however, groups were based on an equal number of projects per group and hence different cost limits for each group, it is still the "small projects" that have the largest average percentage cost overrun with 26% (F=0.666, p=0.576, univariate analysis of variance).

With respect to the net total overrun, very large projects account for the largest overrun; almost half of the total overrun (in amount of overrun) follows from very large projects.

**Project Size as a Predictor for Cost Overruns**

A linear or logarithmic relation between project size and cost overruns is often assumed in literature. The logarithmic relationship provided a slightly better fit for the Dutch data, but since the logarithmic coefficient for the model was not statistically significant (p=0.132), we preferred the simpler, linear model.

Figure 1 shows the plot of percentage cost overruns against project size including the regression line for all projects (solid line) as well as the regression lines for the three project types separately (dotted lines).

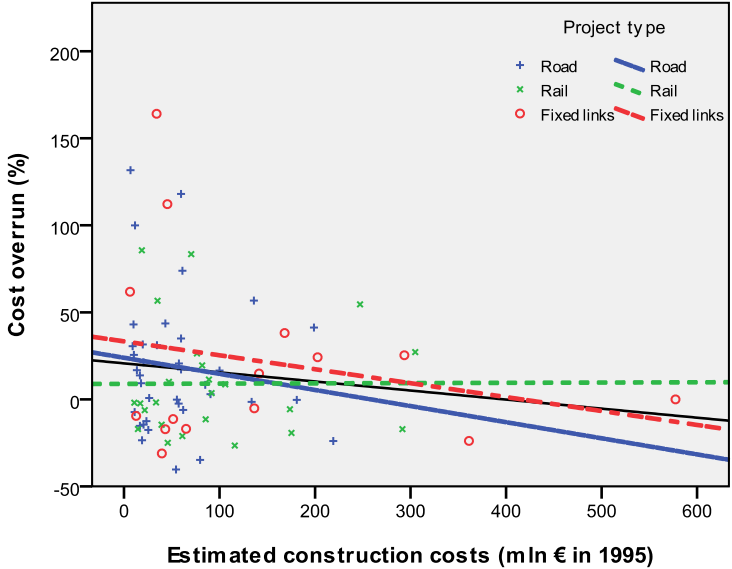

**Figure 1 Estimated costs and cost overruns for 76 projects**

The regression line for cost overrun (%) for all project types is: $\Delta c = 20.61 - 0.03 * c_0$, where $C_0$ is the estimated costs of the project (€ in 1995) Cost overruns decrease with project size; for each additional million Euros that a project costs, the cost overruns decrease by 0.05% (t=-1.083, p=0.282, $R^2$=0.016). Considering the small scope and the low explained variance, we conclude that cost overruns weakly depend on project size.

One fixed link project with considerably higher estimated costs (€ 1006 million) compared to the average (€ 199 million for fixed links project (SD=190)) could be considered a statistical outlier. However, excluding the project from analyses hardly alters the results (slope=-0.036, t=-1.041, p=0.301, $R^2$=0.015).

Considering all the projects together there is no significant relation between project size and the level of cost overruns. However, from figure 1 it can be seen that, especially for fixed link projects, there is a tendency towards smaller cost overruns for larger projects. The relation between the project size and cost overruns was therefore also tested for each project type individually. It turns out that for road and fixed link projects, the same conclusion holds as for all projects; cost overruns decrease with project size. In contrast, for rail projects, cost overruns increase with project size but the effect is negligible (0.001%).

## Implementation Phase

A previous study on cost overruns during project development showed that the main cost overruns occur in the pre-construction phase rather than in the construction phase. This section will therefore also consider the lengths of the pre-construction and construction phase separately as indicators for cost overruns. Note that actual cost overruns can only take place once the project is completed and payments are made, but for simplicity we stick with this term.

### Cost Overruns for Different Lengths of the Implementation Phase

Fixed link projects have the largest average length of the implementation phase (9.2 years, SD=3.2) followed by road projects (7.3 years, SD=3.1) and rail projects (6.5 years, SD=2.3). The length of the implementation phase is statistically different between the three project types (p=0.017). Since fixed link projects also had the largest average cost overrun and rail projects the smallest, we expect a positive relationship between the length of the implementation phase and cost overruns. This was tested in more detail by a regression analysis. Literature assumed either a linear (Flyvbjerg et al. 2004) or quadratic relationship (Odeck, 2004) and both were tested for the Dutch data (there is no reason to assume any other types of non-linear relationships). Although the quadratic relationship resulted in a slightly better fit the coefficient of the quadratic component was not statistically significant (p=0.139) and we therefore preferred the linear relationship.

Figure 2 gives a plot of the cost overrun against the length of the implementation phase for all projects (solid line) and for road, rail and fixed link projects specifically (dotted lines).

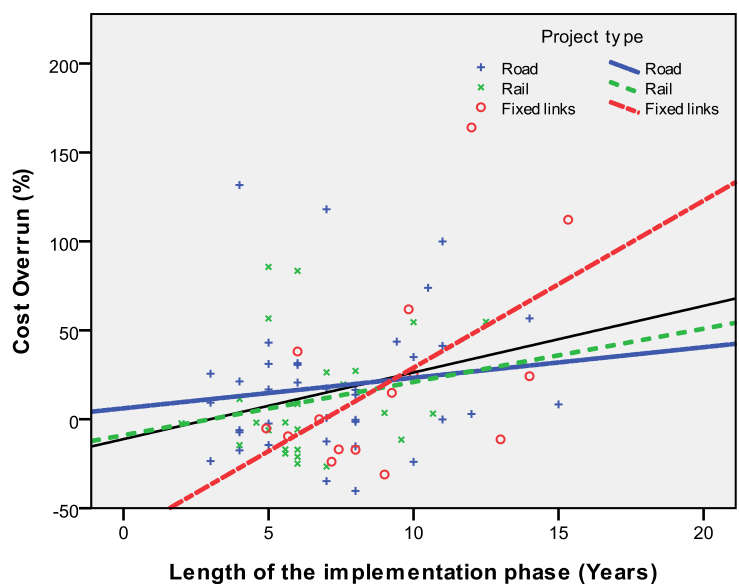

**Figure 2 Length of the implementation phase and cost overruns for 78 projects**

The regression line for all projects is: $\Delta C = -11.15 + 3.74 * T$, where C is the cost overrun (as a % of constant prices) and T is the length of the implementation phase of the project. For each additional year of the implementation phase, cost overruns increase by 3.74% (t=2.533, p=0.013). The explained variance of cost overruns by implementation phase is, however, low ($R^2$ =0.078). For each project type individually, cost overruns also increase with the length of the implementation phase (p<0.05 for road and rail, p=0.036 for fixed link projects).

The implementation phase possibly includes delays. Delay is, at least in the Netherlands, often assumed to be a main predictor of cost overruns. As a delay results in a longer implementation phase and cost overruns increase with each additional year of the implementation phase, it is expected that delays would also influence cost overruns. The average length of the implementation phase for projects with delays is indeed larger (7.7 years) than the average length for projects that were completed on time (6.5 years, t=-1.449, p=0.151, independent sample t-test). The average cost overrun for projects with delays is also larger (18.5% compared to 10.0%) but the difference is not statistically significant (t=-0.787,

p=0.434, independent sample t-test). With a coefficient that is similar to that of the variable length of the implementation phase (3.55, t=1.119, p=0.168) and a lower explained variance of 0.021, it is not the delay but the length of the implementation phase that is the better predictor of cost overruns.

**Cost Overruns for Different Lengths of the Pre-Construction and Construction Phase**

The average length of the pre-construction phase is 3.0 years (SD=2.2) which is significantly shorter than the average length of the construction phase (4.8 years, SD=2.6, t=-3.364, p=0.001, paired sample t-test).

Considering the larger cost overruns in the pre-construction phase compared to the cost overruns in the construction phase, it is expected that the length of the pre-construction phase is more strongly related to cost overruns than the length of the construction phase.

Fixed link projects have the largest pre-construction phase (an average of 3.7 years, SD=2.6), followed by road projects (3.4 years, SD=2.2) and rail projects (1.4 years, SD=0.8, p=0.009). Fixed links also have the largest average cost overruns and we therefore assume a positive relationship between the length of the pre-construction phase and the cost overrun. By means of a regression analysis, this relation can be further examined. Again it was tested for linear and quadratic relationships. The quadratic model has a similar fit to the linear model but since the coefficient of the model is not statistically significant (t=-0.319, p=0.751) the linear model is preferred.

The regression formula is: $\Delta C = -1.25 + 5.02 * T$, where C is the cost overrun (as a % of constant prices) and T is the length of the pre-construction phase of the project. The pre-construction phase is responsible for 10.2% of the variance in cost overruns. For each additional year the pre-construction phase takes, the cost overrun increases by 5.0% (t=2.365, p=0.022).

For road, rail and fixed links, there is also a positive relation between the length of the pre-construction phase and cost overruns, with varying degrees of impacts. The extent by which cost overruns increase with each additional year of the pre-construction phase is much larger for rail projects than for road and fixed link projects.

Rail projects have the largest construction phase, with an average of 5.6 years (SD=2.3) followed by fixed link projects (4.9 years, SD=2.6) and road projects (4.3 years, SD=2.6) (p=0.305, Anova). Similar to the lengths of the implementation and pre-construction phase, the relation between the length of the construction phase and cost overruns is considered by means of a regression analysis. The data was also tested for a quadratic relationship but this did not result in a better model fit. The regression formula is: $\Delta C = 14.42 - 0.15 * T$, where C is the cost overrun (as a % of constant prices) and T is the length of the construction phase of the project. For each additional year of construction, the extent of the cost overrun decreases by 0.15% (t=-0.075, p=0.940). Beside this small coefficient, the small explained variance in cost overruns ($R^2$=0.000) shows that cost overruns are only to a very small extent dependent on the length of the construction phase. The influence of the construction phase on cost overruns is also small for each project type individually. Two projects were identified as statistical outliers, having a construction phase of more than 10 years. Although the relation between length of the construction phase and cost overruns changes into a positive one, the difference is only 1% and not statistically significant (p=0.733).

## Conclusions and Discussion

### Conclusions

This study addressed the influence of the project type, project size and the length of the implementation phase on cost overruns for Dutch projects.

First, the main findings regarding the project type are as follows:

- Rail projects perform better compared with road and fixed link projects.
- Road projects are particularly vulnerable to cost overruns.
- For all project types, cost overruns mainly appear in the pre-construction phase.

Considering these findings it can be concluded that also regarding project type, the cost performance in the Netherlands differs from those worldwide. Rail projects have the largest average cost overrun worldwide, whereas in the Netherlands this is the category with the smallest average overrun. In addition to the lower average cost overrun, the frequency of cost overruns for all project types is considerably lower compared to worldwide findings.

Secondly, for project size the following conclusions can be drawn:

- The problem of cost overruns is most severe for small projects.
- Project size does not significantly influence the cost overrun.

In average percentages, cost overruns are highest for small projects, but the impact of project size on cost overruns is small. However, in terms of net total overrun in percentages, larger projects contribute to a greater extent to cost overruns.

Thirdly, regarding the implementation phase, the main findings are as follows:

- The longer the implementation phase the higher the cost overruns, especially for fixed link projects.
- The pre-construction phase is significantly shorter than the construction phase but it has the highest influence on cost overruns.

The length of the pre-construction phase has a strong (positive) relation and the length of the construction phase has a weak (negative) relation with cost overruns. This makes the length of the pre-construction phase a much better determinant of cost overruns than the length of the construction phase. It is even a better predictor than the length of the implementation phase. The same applies for the individual project types, except for fixed link projects. We therefore conclude that the focus should lie on the pre-construction phase when searching for causes and cures for cost overruns, at least for the Netherlands.

**Discussion**

The findings raise several points for discussion. This section will address these subjects focussing on the possible reasons that can explain the findings.

The study showed that in the Netherlands cost overruns for rail projects are relatively low, both when compared nationally with roads and fixed links and internationally when compared with worldwide findings. The difference between project types may be related to the organisational set-up and institutional settings which is different for rail projects (with ProRail as project owner) and for road projects (with RWS as project owner).

The type of construction, i.e. either new construction or the broadening of an existing structure, does not explain the difference between road and rail projects in the Netherlands, but it could explain the difference with the worldwide findings. In the Dutch data, a large share of projects concerns broadenings, adjustments, improvements and not new infrastructure. New infrastructure typically involves larger cost overruns and the international database may include a greater number of new infrastructures, which could explain the difference. Furthermore, the type of rail, heavy or light rail, could explain the difference with the worldwide findings.

This research furthermore concluded that small projects have the largest average cost overrun. Odeck (2004) suggests that this could be due to the greater amount of attention that is given to larger projects. "Larger projects are most probably under much better management as compared to smaller ones". This suggests that smaller projects deserve more attention than is currently the case as they result in similar cost overruns as the large projects. Of course the benefits should exceed the additional management costs. In addition, the length of the pre-construction phase turns out to be a better predictor of cost overruns than the length of the construction phase or implementation phase. The causality is still uncertain, so it is as yet impossible to conclude that shortening this period will reduce the magnitude of cost overruns. In addition, a shorter phase might not be sufficient to obtain agreement on the project implementation. This would have to be discussed in the construction phase and hence, cost increases might not be reduced but shifted to the next phase. More insight into the reasons behind the cost increase in this phase is needed to determine the effect of shortening the length of the pre-construction phase.  It is questioned whether a shorter pre-construction phase fits with the decision-making culture in the Netherlands. This culture is characterised by many opportunities for the general public, as well as the local citizens, interest groups and industry to participate in the process. The belief is that this will, eventually, result in greater support for the project's plans, therefore avoiding resistance in later phases of the decision-making process. Depending upon the level of participation in this pre-construction phase, reducing the phase might complicate the possibilities for participation and not necessarily result in smaller cost overruns.  Shortening this phase may even result in larger cost overruns because possibilities for participation are reduced that may have corrected the cost estimate.

Considering the three main explanations for cost overruns (Flyvbjerg et al., 2002) – technical, psychological and political-economic explanations – the latter seems the most likely. Technical explanations concern forecasting errors in technical terms such as inaccurate models (Ascher, 1978; Flyvbjerg et al., 2003a; Morris & Hough, 1987; Wachs, 1990). Psychological explanations are based on the cognitive mind of forecasters resulting in optimistic forecasts. According to this explanation, promoters and forecasters are held to be overly optimistic about project outcomes in the appraisal phase, when projects are planned and decided (Fouracre et al., 1990; Mackie & Preston, 1998; Walmsley & Pickett, 1992; World Bank, 1994 in Flyvbjerg et al., 2002), and political-economic explanations are based on strategic misrepresentation (Wachs, 1989).  The forecasting models or optimism bias do not change with the length of the pre-construction phase and hence cannot explain the increasing cost overruns. Conversely, strategic misrepresentation can increase cost overruns.

In the pre-construction phase, strategic misrepresentation can be seen by the many scope changes. Before the decision to build, the estimated costs were purposefully kept low (usually with a small project scope) to increase the chances for the proposal being accepted. Once accepted, attempts are made to add scope to the project, resulting in large cost increases. The longer the pre-construction period will take, the more opportunities there are to adjust the project plans (either due to unforeseen events or purposefully) and hence raise the project costs and eventually cost overruns. Shortening this phase will however only result in lower cost overruns when the possibility of purposeful scope changes and similar behaviour is eliminated.

## Areas for Further Research

The findings of this research pose several areas for further research. First of all, most importantly, it is interesting to examine the reasons behind the findings that are presented in this research. We suggest hereto various in-depth interviews with persons that were involved in the different projects. Particular attention can be paid to the pre-construction phase, why the costs increased to such a large extent and how reducing the length of the phase would affect the cost development. Further, it would be interesting to disaggregate the cost overruns in different types of costs. Data availability did not allow to make such a distinction in the current research and additional data collection is therefore necessary. Secondly, it would be interesting to compare the cost performance between countries. Thirdly, cost overruns could be considered from the perspective of the decision-making culture or more specifically, the system of governance. It is expected that the way in which decisions are made will influence the cost performance of projects. A first possible distinction could be democratic versus non-democratic systems of governance but other distinctions may also be suitable. We will explore these subjects, the comparison between countries and systems of governance in subsequent papers.

This paper concludes by proposing three additional areas of further research. First of all, the relation between the different phases in the decision-making phase and the extent of the cost overruns could be considered and compared between countries. This would provide an answer to the question of whether the length of the pre-construction phase is also a better indicator of the total cost overruns in other countries. For this, the specific decision-making phases for each country should be taken into account, because it probably varies. Secondly, it might be useful to consider several projects in more detail, to determine specifically the reasons for each cost increase. Lastly, as type of construction (new infrastructure or not) may make a difference to cost performance, additional research into this variable is recommended.

## Acknowledgement

The authors thank the Dutch Ministry of Infrastructure and the Environment, including RWS and KiM for their support in data collection. Furthermore, the authors thank various other institutes who provided valuable data for projects. Special thanks to two anonymous referees for their useful comments on an earlier draft of this paper. The research was supported by the Dutch Ministry of Infrastructure and the Environment.

## Notes

1. For a full description of the project selection, data collection and methodology we refer to Cantarelli *et al.* (forthcoming) which is a companion paper to the present paper. The full methodological elucidation will be included in a PhD Thesis (2011), of which this paper makes up a part.
2. The translation of the MIRT in English is based on:
   http://www.verkeerenwaterstaat.nl/english/topics/water/delta_programme/rules_and_fra mework_of_the_mirt (consulted 20-03-2010)

## List of Figures

## List of Tables